\documentclass[12pt]{article} 
\def\cdate{{September 20, 2019}}
\usepackage{amsfonts}
\usepackage{amsmath}
\usepackage{latexsym}
\usepackage{amssymb}
\usepackage{txfonts}
\usepackage[american]{babel}

\makeatletter
\def\timenow{%
\@tempcnta=\time \divide\@tempcnta by 60 \number\@tempcnta:\multiply
\@tempcnta by 60 \@tempcntb=\time \advance\@tempcntb by -\@tempcnta
\ifnum\@tempcntb <10 0\number\@tempcntb\else\number\@tempcntb\fi}

\renewcommand{\@oddhead}{\raisebox{0pt}[\headheight][0pt]{\vbox{\hbox
to \textwidth{\textit{\rightmark}\hfil\strut\thepage}
}}}
\renewcommand{\@evenhead}{\raisebox{0pt}[\headheight][0pt]{\vbox{\hbox
to \textwidth{\thepage\hfil\strut\textit{\leftmark}}
}}}
\renewcommand{\@oddfoot}{
}
\renewcommand{\@evenfoot}{
}

\makeatother
\def\RR{{\mathbb R}}

\def\cD{{\cal D}}

\def\cM{{\cal M}}

\def\cV{{\cal V}}
\def\cW{{\cal W}}
\def\cP{{\cal P}}
\def\cR{\mathcal{ R}} 

\def\tr{\mathrm{ tr\,}} 
\def\Tr{\mathrm{ Tr\,}}

\def\Real{\mathrm{Re\,}}

\def\dim{\mathrm{dim\,}}
 
\def\vol{\mathrm{vol}}

\def\be{\begin{equation}} 
\def\ee{\end{equation}} 
\def\bea{\begin{eqnarray}} 
\def\eea{\end{eqnarray}} 
\def\bed{\begin{definition}{\ }}
\def\eed{\end{definition}}
\def\bd{\begin{description}}
\def\ed{\end{description}}
\def\bc{\begin{center}}
\def\ec{\end{center}}

\newtheorem{theorem}{Theorem}
\newtheorem{lemma}{Lemma}

\newtheorem{definition}{Definition}

\def\sideremark#1{\ifvmode\leavevmode\fi\vadjust{\vbox to0pt{\vss
\hbox to 0pt{\hskip\hsize\hskip1em
\vbox{\hsize2cm\tiny\raggedright\pretolerance10000
\noindent #1\hfill}\hss}\vbox to8pt{\vfil}\vss}}}

\begin{document}

\begin{titlepage}

\null
\vspace{-10mm}
\hspace*{50truemm}{\hrulefill}\par\vskip-4truemm\par
\hspace*{50truemm}{\hrulefill}\par\vskip5mm\par
\hspace*{50truemm}{{\large\sc New Mexico Tech {\rm 
(\cdate)}}}\vskip4mm\par
\hspace*{50truemm}{\hrulefill}\par\vskip-4truemm\par
\hspace*{50truemm}{\hrulefill}
\par
\bigskip
\bigskip
\par
\par
\vspace{3cm}
\centerline{\huge\bf Bogolyubov Invariant}
\bigskip
\centerline{\huge\bf via Relative Spectral Invariants}
\bigskip
\centerline{\huge\bf on Manifolds}
\bigskip
\bigskip
\centerline{\Large\bf Ivan G. Avramidi
}
\bigskip
\centerline{\it Department of Mathematics}
\centerline{\it New Mexico Institute of Mining and Technology}
\centerline{\it Socorro, NM 87801, USA}
\centerline{\it E-mail: ivan.avramidi@nmt.edu}
\bigskip
\medskip
\vfill
{\narrower
\par
We introduce and study { new} spectral invariant
of {two} elliptic partial differential operators of Laplace 
and Dirac type on compact smooth manifolds without boundary that 
depends on both the eigenvalues and the eigensections
of the operators, which 
is a equal to the regularized number of created 
particles from the vacuum when the dynamical operator 
depends on time. We study the asymptotic expansion of this invariant
for small adiabatic parameter and compute explicitly the
first two coefficients of the asymptotic expansion.

\par}

\vfill

\end{titlepage}


 
\section{Introduction}
\setcounter{equation}0


We introduce and study a new relative spectral invariant of 
two elliptic operators.
We hope that these new invariant could shed new light on the 
old questions of spectral
geometry: 
``Does the spectral information of {\it two} elliptic operators determine
the geometry of a manifold?''
This invariant appears naturally in the study of particle creation in quantum field
theory and quantum gravity \cite{birrel80,dewitt75}. It is equal to the number of 
created particles from the vacuum when the dynamical operator depends on time. 
That is why, we will just call it
the {\it Bogolyubov invariant}.
This is a continuation of our work on the systematic study
of generalized spectral invariants 
initiated in 
\cite{avramidi16, avramidi17} where we studied 
heat determinant and quantum heat traces
and, in particular, \cite{avramidi19a}, where we studied
relative spectral invariants.

In Sec. \ref{sec2} we consider an idealized case of
an instantenous change of the operators and express the 
Bogolyubov invariant formally in terms of the heat traces
and relative spectral invariants.
In Sec. \ref{sec3} we introduce a rigorous definition
of the Bogolyubov invariant as a trace of 
some non-trivial functions of elliptic operators
and describe some of its properties.
In Sec \ref{sec4} we derive some reduction formulas
that express the Bogolyubov invariant in terms of the 
combined heat traces. 
In Sec. \ref{sec5} we establish the asymptotics of the
combined heat traces.
In Sec. \ref{sec6} we use Mellin transform to compute the 
asymptotics of the Bogolyubov invariant
and explicitly compute the first two coefficients.
In Sec. \ref{sec7} we consider some examples.

\subsection{Particle Creation in Quantum Field Theory}
\setcounter{equation}0

To motivate the introduction of this invariant
we describe now the standard method for calculation of particles
creation via the Bogolyubov transformation. Let $({\cal M}, h)$ be a
pseudo-Riemannian $(n+1)$-dimensional 
spin manifold with a Lorentzian metric $h$. 
We assume that $({\cal M}, h)$ is
globally hyperbolic so that there is a foliation of ${\cal M}$ with space
slices $M_t$ at a time $t$, moreover, we assume that there is a global time
coordinate $t$ varying from $-\infty$ to $+\infty$ and that at all times
$M_t$ is a compact $n$-dimensional Riemannian manifold without
boundary. We will also assume that there are well defined limits
$M_{\pm}$ as $ t\to \pm\infty$. For simplicity, we will just assume that
the manifold ${\cal M}$ has two cylindrical ends, $(-\infty,\beta)\times M$ and
$(\beta,\infty)\times M$ for some positive parameter $\beta$. So, the foliation
slices $M_t$ depend on $t$ only on a compact interval $[-\beta,\beta]$.
Let $g_t$ be  the induced  Riemannian metric on $M_t$ and
$d\vol_{g_t}=(\det g^t_{ij})^{1/2}dx$ be the corresponding Riemannian volume element.
We label the space-time
indices that run over $(0,1,\dots,n)$ by Greek letters and the space indices 
that run over $(1,\dots,n)$ by Latin letters. 
Let $\cW$ be a {\it real}
vector bundle over $\cM$ and $\cV_t$ be the corresponding time
slices (vector bundles over $M_t$) (any complex vector bundle can be made real
by just doubling the dimension). Henceforth we will omit the index $t$ on
$M$, $g$ and $\cV$ when it does not cause a confusion. 
We define the natural $L^2$ inner product 
\be
\left(\varphi_1,\varphi_2\right)_{M}
=\int\limits_{M} d\vol_g\; \left<\varphi_1,\varphi_2\right>\,,
\ee
where $\left<\;,\;\right>$ is the fiber product in $\cV$,
and the space $L^2(\cV)$  of square integrable sections of the bundle $\cV$.

In quantum field theory there are two types of particles, bosons and fermions.
The bosonic fields are described by second order Laplace type partial differential operators
whereas the fermionic fields are described by first order Dirac type partial differential 
operators.

\subsection{Bosonic Fields}

Bosonic fields are described by sections of tensor bundles
(or more generally by twisted tensor bundles); so, let $\cW$ 
be a tensor bundle.
Let $H_t$ be a
one-parameter family of {\it self-adjoint elliptic
second-order} partial differential operators acting on smooth sections of
the bundle $\cV_t$. 
We assume that there are well defined limits $H_{\pm}$ as $t\to
\pm\infty$, that is, $H_{\pm}$ are elliptic self-adjoint
second-order partial differential operators acting on sections of $\cV$
over $M$. 

Let $m$ be a sufficiently large positive parameter
so that the operator $H+m^2$ is positive and the
following  {\it pseudo-differential} operators
\be
\omega_t=(H_t+m^2)^{1/2}
\ee
are well defined.
Henceforth, we will omit the index $t$ on all operators.

We define the hyperbolic operator for bosonic fields 
by
\be
L=\partial_t^2+H\,.
\ee
Then the dynamics of the bosonic fields is described by the 
space ${\cD}$ of solutions of the hyperbolic equation
\be
(L+m^2)\varphi=0\,
\label{226xx}
\ee
with the inner product
\be
(\varphi_1,\varphi_2)_{\cD}
=\left(\varphi_1,i\partial_t\varphi_2\right)_M
+\left(i\partial_t\varphi_1,\varphi_2\right)_M\,.
\label{24viax}
\ee
It is easy to show that this inner product is well  defined since it
does not  depend on the time $t$.

\subsection{Fermionic Fields}

The fermionic fields are described by sections of spin-tensor bundles
(or, more generally, by twisted spin-tensor bundles);
so, let $\cW$ be a spin-tensor bundle.
Let $\xi$ be a self-adjoint involutive endomorphism of the bundle ${\cal V}$,
\be
\xi^2=I, \qquad \xi^*=\xi,
\ee
with $I$ denoting the identity endomorphism,
defining a
natural decomposition of the vector bundle $\cV=\cV_{(+)}\oplus\cV_{(-)}$, so that
$\dim\cV_{(+)}=\dim\cV_{(-)}$ and 
in the canonical basis
\be
\xi=\left(
\begin{array}{cc}
I & 0\\
0 & -I \\
\end{array}
\right).
\ee
Let $\eta$ be another  self-adjoint involutive endomorphism of the bundle 
${\cal V}$ that anti-commutes with $\xi$,
\be
\eta^2=I, \qquad \eta^*=\eta,
\ee
\be
\xi\eta=-\eta\xi,
\ee
 and has the form
(in the same basis)
\be
\eta=\left(
\begin{array}{cc}
0 & I\\
I & 0 \\
\end{array}
\right).
\ee
This enables one to define an anti-self-adjoint anti-involution
\be
C=\xi\eta=\left(
\begin{array}{cc}
0 & I\\
-I & 0 \\
\end{array}
\right),
\ee
that anti-commutes with both $\xi$ and $\eta$.
(Of course, $\xi,\eta$ and $iC$ are nothing but Pauli matrices.)

Let $A_t$ be a  one-parameter family of
{\it self-adjoint first-order elliptic} partial differential
operators acting on sections of the bundles $\cV_{(\pm)}$ such that its square
\be
H_t=A_t^2
\ee
is a self-adjoint second-order elliptic partial differential operator. 
We assume that there are well defined limits
$A_{\pm}$ as $t\to
\pm\infty$, that is, $A_{\pm}$ are elliptic self-adjoint
first-order partial differential operators acting on sections of $\cV_{(\pm)}$
over $M$. As mentioned above we drop the subscript $t$ on all operators
below.

Let $K$ be a {\it self-adjoint} 
elliptic {first-order} partial differential operator
acting on smooth sections of the bundle $\cV$ defined by
\be
K=\xi\otimes A
=\left(
\begin{array}{cc}
A & 0\\
0 & -A \\
\end{array}
\right)\,.
\label{211ccx}
\ee
Note that the operator $K$ anti-commutes with $\eta, C$ and commutes with $\xi$
\be
K\eta=-\eta K, \qquad
KC=-CK, \qquad
K\xi=\xi K.
\ee
Therefore, the operator
\be
K+m\eta
=\left(
\begin{array}{cc}
A & m\\
m & -A \\
\end{array}
\right),
\label{214xxa}
\ee
with some mass parameter $m$,
is also a self-adjoint eliptic first-order operator.

The square of the operator $K$
\be
K^2=
\left(
\begin{array}{cc}
H & 0\\
0 & H \\
\end{array}
\right)\,
\ee
is a self-adjoint second-order operator,
and, therefore, the operator
\be
(K+m\eta)^2
=\left(
\begin{array}{cc}
H+m^2 & 0\\
0 & H+m^2 \\
\end{array}
\right)
\label{214xxx}
\ee
is also self-adjoint and {\it positive} for any non-zero 
mass parameter, $m>0$.
We will find it useful to define also 
the {\it pseudo-differential} operator
\be
\omega =|K+m\eta| = (H+m^2)^{1/2}.
\ee

It is easy to see that the operator $(K+m\eta)$ anti-commutes
with the endomorphism $C$,
\be
(K+m\eta)C=-C(K+m\eta),
\label{216viax}
\ee
and, therefore, its spectrum (which is obviously real and non-zero)
 is symmetric with respect to zero, since for every eigensection
$\varphi$ with an eigenvalue $\lambda$ there is
an eigensection $C\varphi$ with the opposite eigenvalue $(-\lambda)$.
Note, though, that it does not mean that the spectrum of the operator $A$
is symmetric with respect to the origin. 
That is why, it is useful to define the spectral projections 
on the positive part of the spectrum
by
\be
P_{}=\frac{1}{2}\left(I + F\right)\,,
\ee
where
\be
F =(K+m\eta) \omega^{-1}
=\left(
\begin{array}{cc}
A\omega^{-1} & m\omega^{-1}\\
m\omega^{-1} & -A\omega^{-1} \\
\end{array}
\right)\,,
\ee
is an involution,
\be
F^2=I.
\ee
Obviously, the operator
\be
I-P=\frac{1}{2}\left(I - F\right)
\ee
is the projection on the negative part of the spectrum.

We define the dynamical hyperbolic operator for the fermionic fields by
\be
D=i\partial_t + K;
\ee
then
\be
-(\eta D)^2
=\left(
\begin{array}{cc}
\partial_t^2+H & 0\\
0 & \partial_t^2+H \\
\end{array}
\right).
\ee
The dynamical space $\cD$ is now the space of solutions of the hyperbolic 
equation
\be
(D+m\eta)\varphi=0\,
\label{226xxa}
\ee
with the inner product 
\be
(\varphi_1,\varphi_2)_{\cal D}
=\left(\varphi_1,\varphi_2\right)_M\,.
\ee
This inner product is well defined since it does not depend on
the time.
Notice also that 
\be
(\eta D+m)(-\eta D+m)
=\left(
\begin{array}{cc}
\partial_t^2+H+m^2 & 0\\
0 & \partial_t^2+H+m^2 \\
\end{array}
\right).
\ee


\subsection{Particle Creation}

Now, we introduce two different bases in the dynamical space $\cD$
(the space of solutions of the dynamical equations 
(\ref{226xx}) and (\ref{226xxa})).
First, we introduce the so-called {\it positive-frequency in-modes},
$\{\varphi_k\}_{k=1}^\infty$,
by requiring them to satisfy the asymptotic conditions
\be
\lim_{t\to -\infty}(\partial_t +i\omega^-_k)\varphi_k=0\,
\ee
for some positive constants $\omega^{-}_k$, and normalize them by
\be
(\varphi_i,\varphi_j)_{\cD}=\delta_{ij}\,.
\ee
We also introduce another basis, so called
{\it negative-frequency out-modes},
$\{\psi_k\}_{k=1}^\infty$,
by requiring them to satisfy the asymptotic conditions
\be
\lim_{t\to +\infty}(\partial_t-i\omega^+_k)\psi_k=0\,
\ee
for some positive constants $\omega^{+}_k$, and normalize them by
\be
(\psi_i,\psi_j)_{\cD}
=-\delta_{ij}\,,
\ee
for bosons
and
\be
(\psi_i,\psi_j)_{\cD}
=\delta_{ij}\,
\ee
for fermions.


Then the total number of created in-particles in the out-vacuum  is 
determined by the so-called {\it Bogolyubov coefficients} and is equal 
(in
both bosonic and fermionic cases) to
\cite{dewitt75,birrel80,grib94,wald94}
\be 
N=\sum_{i,j=1}^\infty|(\varphi_i,\psi_j)_{\cD}|^2.
\ee
Although, it is impossible to calculate this invariant in the general case,
it is easy to see that it is 
always non-negative and vanishes if the operators
 $H$ and $K$
do not depend on time at all. 

\section{Instantaneous Jumps}
\setcounter{equation}0
\label{sec2}

We will consider two limiting cases: i) $\beta\to \infty$, which corresponds
to a slow ({\it adiabatic}) variation of the operators $H_t$ and $K_t$,
and ii) $\beta\to 0$, which corresponds to an 
{\it instantaneous} change of the operators.
In the limit $\beta\to 0$ 
one can compute the
total number of created particles, at least formally. 

\subsection{Bosonic Fields}

We consider the bosonic case first.
Let $\{\omega^\pm_k\}_{k=1}^\infty$ be the eigenvalues (counted with multiplicities)
and $\{u^\pm_k\}_{k=1}^\infty$ be the corresponding orthonormal
sequence of the eigensections of the operators $\omega_\pm$;
obviously, all eigenvalues are positive, $\omega^\pm_k>0$.
Also, let $P^\pm_k$ be the orthogonal projections to the eigenvectors
$u^\pm_k$.
Here every eigenvalue and eigensection is taken with its multiplicity.
We assume that the dynamical modes are continuously 
differentiable in time, that is, they
are continuous in time
and have continuous first partial time derivative.
Then the positive-frequency in-modes and the negative frequency out-modes
are
\be
\varphi^{}_k=
\left\{
\begin{array}{ll}
\displaystyle
\frac{1}{\sqrt{2\omega^-_k}}e^{-i\omega^-_k t}u^-_k, & \mbox{ for } t<0,
\\[15pt]
\displaystyle
\sum_{j=1}^\infty \frac{(u^+_j,u^-_k)}{2\omega^+_j\sqrt{2\omega^-_k}}
\left\{
(\omega^+_j+\omega^-_k)e^{-i\omega^+_j t}
+(\omega^+_j-\omega^-_k)e^{i\omega^+_j t}
\right\}u^+_j, & \mbox{ for } t>0,
\\
\end{array}
\right.
\ee
\be
\psi^{}_k=
\left\{
\begin{array}{ll}
\displaystyle
\sum_{j=1}^\infty \frac{(u^-_j,u^+_k)}{2\omega^-_j\sqrt{2\omega^+_k}}
\left\{
(\omega^-_j+\omega^+_k)e^{i\omega^-_j t}
+(\omega^-_j-\omega^+_k)e^{-i\omega^-_j t}
\right\}u^-_j, & \mbox{ for } t<0,
\\[15pt]
\displaystyle
\frac{1}{\sqrt{2\omega^+_k}}e^{i\omega^+_k t}u^+_k,
& \mbox{ for } t>0.
\\
\end{array}\,
\right.
\ee

By using these modes it is easy to 
show that the total number of created particles in the 
bosonic case is given by a {\it formal} series
\be
N_b = \frac{1}{4}\sum_{j,k=1}^\infty 
\left(\sqrt{\frac{\omega^-_j}{\omega^+_k}}
-\sqrt{\frac{\omega^+_k}{\omega^-_j}}\right)^2 \Tr P^-_jP^+_k;
\ee
here we used the obvious relation
\be
\Tr P^-_jP^+_k =|(u_j^-,u^+_k)|^2.
\ee
This can be written in the form
of a {\it formal} trace
\bea
N_b &=&
\frac{1}{4}\Tr
\left(\omega_--\omega_+\right)
\left(\frac{1}{\omega_+}-\frac{1}{\omega_-}\right)\,.
\label{214aaa}
\eea
Note that this invariant is non-negative, as it should be, $N_b\ge 0$.

However, since the operators $\omega_\pm$ are unbounded
this combination is not a trace class operator;
it diverges at the high-end of the spectrum.
Another way to see this is as follows.
The operators $\omega$ and its inverse $\omega^{-1}$
can be represented in terms of the heat kernel of its square
$\omega^2=H+m^2$
by
\bea
\omega^{-1} &=& \frac{1}{\sqrt{\pi}}\int\limits_0^\infty dt\; t^{-1/2}
e^{-tm^2}
\exp(-tH),
\label{27xxa}
\\
\omega &=& -\frac{1}{\sqrt{\pi}}\int\limits_0^\infty dt\; t^{-1/2}
\partial_t\left[e^{-tm^2}\exp(-tH)\right].
\label{27zza}
\eea
By using these equations we obtain a 
{\it formal} expression of the number of particles in terms of the heat kernels
\be
N_b =
\frac{1}{4\pi}
\int\limits_0^\infty dt \int\limits_0^\infty ds\; s^{-1/2}t^{-1/2}
\partial_t\left\{
e^{-(t+s)m^2}\Psi(t,s)\right\},
\label{238xxa}
\ee
where
\be
\Psi(t,s)=
\Tr\left\{\exp(-tH_+)-\exp(-tH_-)\right\}
\left\{\exp(-sH_+)-\exp(-sH_-)\right\};
\label{220xxa}
\ee
we will call this function a {\it relative heat trace}.
The function $\Psi(t,s)$ is symmetric and can be expressed
further in terms of the traces
\be
\Psi(t,s)=\Theta_+(t+s)+\Theta_-(t+s)
-X(t,s)-X(s,t),
\ee
where
\be
\Theta_\pm(t) =\Tr\exp(-tH_\pm),
\label{237ccx}
\ee
is the {\it classical heat trace} of a Laplace type operator and
\be
X(t,s) =\Tr\exp(-tH_+)\exp(-s H_-)
\label{238ccx}
\ee
is a {\it combined heat trace}. 
For more details, see \cite{avramidi19a}.

It is easy to show that the integral (\ref{238xxa})
converges for large $t$ and $s$ for large positive real $m$;
however, it diverges as $t,s\to 0$. Of course, this
happens because of the idealized situation with the instantaneous jump of the operator
$H$.
For a smooth deformation of the operators
the number of particles is finite. 
The {Bogolyubov invariant} $B_b(\beta)$
we define below is a {\it regularized version of 
the number of particles} that is well defined for any $\beta$ and has the asymptotics
(\ref{214aaa}) and (\ref{238xxa}) as $\beta\to 0$.


\subsection{Fermionic Fields}

As we mentioned above the spectrum of the operators $A^\pm$ is not necessarily
symmetric with respect to zero. Let the eigenvalues 
(counted with multiplicities)
of the operator
$A^\pm$ be $\{\mu^\pm_k\}_{k=1}^\infty$ (both positive and negative;
they can be ordered in their absolute value). 
Then the eigenvalues of the operator
$(K+m\eta)$ are $\{\omega^\pm_k,-\omega^\pm_k\}_{k=1}^\infty$
with
$\omega^\pm_{k}=\left((\mu^\pm)_k^2+m^2\right)^{1/2}$.
Let  and 
$\{u^\pm_{k}, Cu^\pm_{k},\}_{k=1}^\infty$
be the corresponding orthonormal eigensections of the operators
$(K_\pm+m\eta)$. Also, let $P^\pm_k$ and $(-CP^\pm_kC)$ be the
orthogonal projections to the eigensections $u^\pm_k$ and $Cu^\pm_k$.

We require our modes to be continuous in time.
Then the positive-frequency in-modes and negative frequency out-modes
are
\be
\varphi_k=
\left\{
\begin{array}{ll}
\displaystyle
Cu^-_{k}e^{-i\omega^-_k t}, & \mbox{ for } t<0,
\\[15pt]
\displaystyle
\sum_{j=1}^\infty \left\{(u^+_{j}, Cu^-_{k}) u^+_{j} e^{-i\omega^+_j t}
+(Cu^+_{j}, Cu^-_{k}) Cu^+_{j} e^{i\omega^+_j t}
\right\}, & \mbox{ for } t>0,
\\
\end{array}
\right.
\ee
\be
\psi_k=
\left\{
\begin{array}{ll}
\displaystyle
\sum_{j=1}^\infty \left\{(u^-_{j}, u^+_{k}) u^-_{j} e^{-i\omega^-_j t}
+(Cu^-_{j}, u^+_{k}) Cu^-_{j} e^{i\omega^-_j t}
\right\}, & \mbox{ for } t<0,
\\[15pt]
\displaystyle
u^+_{k} e^{i\omega^+_k t}, 
& \mbox{ for } t>0.
\\
\end{array}
\right.
\ee
By using these modes it is easy to show that 
the total number of created fermionic particles
is given by the formal series
\be
N_f =\sum_{j,k=1}^\infty \Tr (-CP^-_jCP^+_k),
\ee
where we used the equation 
$\Tr (-CP^-_jCP^+_k)=|(Cu^-_{j}, u^+_{k})|^2$.
This can be written in the form of a formal trace
\bea
N_f &=& \Tr (I-P^-) P^+ 
=\frac{1}{8}
\Tr (F_+-F_-)^2
\,.
\label{finst}
\eea

%
%

By using the same heat kernel trick we represent the operators $F_\pm$
as
\be
F_\pm=\pi^{-1/2}\int\limits_0^\infty dt\; t^{-1/2}e^{-tm^2}
(K_\pm+m\eta )\exp(-tH_\pm)\,
\ee
to rewrite this {\it formally} in the form
\bea
N_f &=&
\frac{1}{8\pi}
\int\limits_0^\infty dt \int\limits_0^\infty ds\; s^{-1/2}t^{-1/2}
e^{-(t+s)m^2}
\nonumber\\[5pt]
&&\times
\Tr\left\{ (K_++m\eta)\exp(-tH_+)- (K_-+m\eta)\exp(-tH_-)\right\}
\nonumber\\[5pt]
&&\times
\left\{ (K_++m\eta)\exp(-sH_+)- (K_-+m\eta)\exp(-sH_-)\right\}
\label{220xxb}.
\eea
Finally, by using the form (\ref{214xxa})
of the operator $K_\pm$
the finite-dimensional trace can be computed to get
\bea
N_f &=&
\frac{1}{4\pi}
\int\limits_0^\infty dt \int\limits_0^\infty ds\; s^{-1/2}t^{-1/2}
e^{-(t+s)m^2}
\left\{\Phi(t,s)
+m^2\Psi(t,s)
\right\},
\label{221xxb}
\eea
where
\be
\Phi(t,s)=
\Tr\left[A_+\exp(-tH_+)-A_-\exp(-tH_-)\right]
\left[A_+\exp(-sH_+)-A_-\exp(-sH_-)\right].
\label{248xxa}
\ee
is the {\it relative heat trace for Dirac operators}.
This function can be written in the form
\be
\Phi(t,s)=-\partial_t\Theta_+(t+s)-\partial_t\Theta_-(t+s)
-Y(t,s)-Y(s,t),
\ee
where $\Theta_\pm(t)$ are the classical heat traces given by (\ref{237ccx}) and
\be
Y(t,s)=\Tr A_-A_+\exp(-tH_+)\exp(-sH_-)
\ee
is the {\it combined heat trace for the Dirac operators}.

Again, it is easy to show that the integrals  (\ref{220xxb}),
(\ref{221xxb})
converge for large $t$ and $s$ for large positive real $m$
but diverge as $t,s\to 0$. 
This
happens because of the idealized situation with the instantaneous jump of the operators
$K$.
For a smooth deformation of the operators
the number of fermionic particles is also finite. 
We will define below an invariant, 
called {\it Bogolyubov invariant}, $B_f(\beta)$, that is a {\it regularized version} of 
the number of particles that is well defined for any $\beta$ and has the asymptotics
(\ref{finst}) and (\ref{221xxb}) as $\beta\to 0$.


\section{Bogolyubov Invariant}
\setcounter{equation}0
\label{sec3}
\subsection{Laplace Type Operators}

Let $H_\pm$ be two elliptic second-order self-adjoint partial differential operators
with positive definite leading symbol of Laplace type 
and 
$
\omega_\pm=(H_\pm+m^2)^{1/2}.
$
Our goal is to generalize the total number of created particles given by
the trace (\ref{214aaa}) so that it converges
at high-end of the spectrum. We replace $\omega$ by 
 $f(\beta\omega)$ where $f$
is a smooth function 
that is linear as $x\to 0$ and is constant up to 
exponentially small terms as $x\to \infty$
and $\beta$ is a positive real parameter; so we define
\bea
B_b(\beta)
&=&
\frac{1}{4}\Tr
\Biggl\{f(\beta\omega_-)-f(\beta\omega_+)\Biggr\}
\left\{\frac{1}{f(\beta\omega_+)}
-\frac{1}{f(\beta\omega_-)}\right\}.
\label{31nnn}
\eea

In particular, we will use the function
\be
f(x)=\tanh\left(\frac{x}{2}\right).
\ee
By using the equations
\bea
f(x) &=& 1-2E_f(x),\\
{}\frac{1}{f(x)}&=& 1+2E_b(x),
\eea
where
\bea
E_f(x) &=& \frac{1}{e^x+1},
\\ 
E_b(x) &=& \frac{1}{e^x-1},
\eea
we can rewrite the invariant (\ref{31nnn}) in the form
\bea
B_b(\beta)&=&\Tr
\Bigl\{E_f(\beta\omega_+)-E_f(\beta\omega_-)\Bigr\}
\Bigl\{E_b(\beta\omega_+)-E_b(\beta\omega_-)\Bigr\},
\label{311xxa}
\eea
and, further, by using
\be
E_f(x)E_b(x)=E_b(2x),
\ee
in the form
\be
B_b(\beta)=\Tr
\Bigl\{E_b(2\beta\omega_+)+E_b(2\beta\omega_-)
-E_b(\beta\omega_+)E_f(\beta\omega_-)
-E_b(\beta\omega_-)E_f(\beta\omega_+)
\Bigr\}.
\ee
Notice that this invariant is closely related to the 
{\it quantum heat traces}
\be
\Theta_{f,b}(\beta,\mu)=\Tr E_{f,b}\left[\beta(\omega-\mu)\right]
\ee
studied in \cite{avramidi17}.

Let $(\lambda_k^\pm)_{k=1}^\infty$ and $(P_k^\pm)_{k=1}^\infty$ be the 
eigenvalues and the 
corresponding projections to the eigenspaces of the operators $H_\pm$;
then $\omega^\pm_k=(\lambda^\pm_k+m^2)^{1/2}$
are the eigenvalues of the operators 
$\omega_\pm$.
We order the eigenvalues so that they form an increasing sequence of real numbers.
In a special case when the operators $H_-$ and $H_+$ have the same
projections, $P_k^-=P_k^+$,
but different eigenvalues this invariant takes the form
\be
B_b(\beta)=\sum_{k=1}^\infty
\frac{\sinh^2[\beta(\omega_k^+-\omega_k^-)/2]}{\sinh(\beta\omega_k^+)
\sinh(\beta\omega_k^-)}.
\ee
%
%

Invariants like this come up in the study of creation of bosonic particles
in quantum field theory and quantum gravity 
in external classical fields
\cite{birrel80}. It is equal to the number of 
created particles from the vacuum when the dynamical operator is changed in time
from $H_-$ at $t\to -\infty$ to $H_+$ at $t\to+\infty$.
In fact, this is equal {\it exactly}
to the number of created particles in a cosmological model
considered in \cite{birrel80}.
 That is why, we will 
just call it
the {\it Bogolyubov invariant}.

It is easy to compute the behavior of this invariant for the asymptotic cases
$\beta\to\infty$ and $\beta\to 0$.
In the adiabatic limit $\beta\to \infty$ the invariant $B_b$ takes the form
\bea
B_b(\beta)&\sim&
\Tr\Biggl\{
e^{-2\beta\omega_-}
+e^{-2\beta\omega_+}
-2e^{-\beta\omega_-}
e^{-\beta\omega_+}
\Biggr\}\,.
\eea
This expression is closely related to the 
{\it relativistic
heat trace} 
\be
\Theta_r(\beta)=\Tr \exp(-\beta\omega)
\ee
studied in \cite{avramidi17}.
The limit $\beta\to 0$ is singular; {\it formally}, we obtain
\bea
B_b(\beta) &\sim&
\frac{1}{4}\Tr
\Biggl(\omega_--\omega_+\Biggr)
\left(\frac{1}{\omega_+}-\frac{1}{\omega_-}\right)\,,
\eea
which is similar to (\ref{214aaa}).
That is why, the invariant $B_b(\beta)$ is the generalization of the number of
created bosonic particles $N_b$. However, this trace diverges.

Let $g_\pm^{ij}$ be the metrics determined by the leading symbols
of the Laplace type operators $H^\pm$, i.e.
\be
\sigma_L(H_\pm;x,\xi)=I g_\pm^{ij}(x)\xi_i\xi_j=I|\xi_\pm|^2.
\ee
By using the results of our paper 
\cite{avramidi17} (or by using the calculus of pseudodifferential operators)
 one can show that as $\beta\to 0$
\be
B_b(\beta) = \beta^{-n}V_b +O(\beta^{-n+2}),
\ee
where $V_b$ is an invariant (depending only on the metrics $g_+$ and $g_-$)
defined by
\bea
V_b=N\int_M dx\;\int_{\RR^n}\frac{d\xi}{(2\pi)^n}
\frac{\sinh^2\left[\left(|\xi_+|-|\xi_-|\right)/2\right]}{\sinh(|\xi_+|)\sinh(|\xi_-|)}
\label{417viax}
\eea
with $N=\tr I=\dim \cV$. 
It is easy to see that this integral converges.
We will compute it
in an alternative form below.

\subsection{Dirac Type Operators}

Let  $K_\pm$ be elliptic first-order self-adjoint partial differential operators
of Dirac type described above, (\ref{211ccx}),
$\omega_\pm=|K_\pm+m\eta|=(H_\pm+m^2)^{1/2}$, and
$F_\pm=(K_\pm+m\eta)\omega_\pm^{-1}$.
Our goal is to generalize the trace (\ref{finst}) so that it converges
at high-end of the spectrum. We replace the operators $F_\pm$ by the 
operators $F_\pm g(\beta\omega_\pm)$ where $g$ is a smooth function 
such that $g(x)\sim 1$ as $x\to 0$ and 
is exponentially small as $x\to\infty$.
That is, we define the fermionic Bogolyubov invariant by
the trace of a trace class operator
\bea
B_f(\beta) &=& \frac{1}{8}\Tr
\Bigl\{F_+g(\beta\omega_+)
-F_-g(\beta\omega_-)\Bigr\}^2.
\eea
We choose the function $g$ in the form
\be
g(x)=\frac{x}{\sinh x}.
\ee
Let $E_0$ be a function defined by
\be
E_0(x)=\frac{1}{2}(E_b(x)+E_f(x))=\frac{1}{2\sinh x};
\ee
Then the Bogolyubov invariant has the form
\be
B_f(\beta)=\frac{\beta^2}{2}
\Tr\Bigl\{(K_++m\eta) E_0(\beta\omega_+)
-(K_-+m\eta) E_0(\beta\omega_-)\Bigr\}^2.
\ee
By computing the final-dimensional trace it can be reduced to 
\bea
B_f(\beta) &=& \beta^2
\Tr\Biggl\{\left[A_+E_0(\beta\omega_+)-A_-E_0(\beta\omega_-)\right]^2
+m^2\left[E_0(\beta\omega_+)-E_0(\beta\omega_-)\right]^2
\Biggr\}
\nonumber
\\
&=& \beta^2\Tr
\Biggl\{\omega_+^2E^2_0(\beta\omega_+)
+\omega_-^2E^2_0(\beta\omega_-)
-2(A_+A_-+m^2)E_0(\beta\omega_-)E_0(\beta\omega_+)
\Biggr\}
\nonumber\\
\label{318xxa}
\eea

In the adiabatic limit, as $\beta\to \infty$, the invariant $B_f$ takes the form
\bea
B_f(\beta) &\sim&
\beta^2\Tr
\Bigl\{\omega^2_+ e^{-2\beta\omega_+}+\omega^2_-e^{-2\beta\omega_-}
-2(A_+A_-+m^2)e^{-\beta\omega_-}e^{-\beta\omega_+}
\Bigr\}.
\eea
As in the bosonic case the limit $\beta\to 0$ is singular;
 formally we get a divergent trace
\bea
B_f(\beta)&\sim&
\frac{1}{8}\Tr
\left(F_+-F_-\right)^2\,,
\eea
similar to (\ref{finst}). Therefore, the invariant $B_f(\beta)$ is the generalization of the
number of created fermionic particles $N_f$.
Obviously, this trace also diverges.

Let $\gamma_\pm^j$ be the Dirac matrices determined by the leading symbol of he
Dirac type operators $A_\pm$, i.e.
\be
\sigma_l(A_\pm;x,\xi)=-\gamma_\pm^j(x)\xi_j.
\label{425viax}
\ee
Then
\be
\gamma_\pm^j(x)=e^j_{\pm,a}(x)\gamma^a,
\ee
where $e^j_{\pm,a}$ are the orthonormal bases for the metrics
$g^{ij}_\pm$,
\be
g^{ij}_{\pm}=\delta^{ab}e^j_{\pm,a}e^j_{\pm,b}.
\ee
By using the results of our paper 
\cite{avramidi17} (or by using the calculus of pseudodifferential operators)
one can show that the leading asymptotics as $\beta\to 0$ is
\be
B_f(\beta) = 
\beta^{-n}V_f +O(\beta^{-n+2}),
\ee
where $V_f$ is an invariant defined by
\bea
V_f=\frac{N}{4}\int\limits_M dx\int\limits_{\RR^n}\frac{d\xi}{(2\pi)^n}
\left\{\frac{|\xi_+|^2}{\sinh^2(|\xi_+|)}
+\frac{|\xi_-|^2}{\sinh^2(|\xi_-|)}
-2\frac{|\xi_+\xi_-|}{\sinh(|\xi_+|)\sinh(|\xi_-|)}
\right\}, 
\label{429viax}
\eea
with $N=\tr I=\dim \cV_{(\pm)}$ and  
\be
|\xi_+\xi_-|=
\frac{1}{2}\left(\gamma_+^{i}\gamma_-^{j}
+\gamma_-^{j}\gamma_+^{i}
\right)\xi_i\xi_j
=\delta^{ab}e_{+,a}^{i}e_{-,b}^{j}\xi_i\xi_j.
\ee
This integral obviously converges. We will compute it in an alternative form
later.

\section{Heat Traces}
\setcounter{equation}0
\label{sec4}
\subsection{Reduction Formulas}

Let $H$ be a positive self-adjoint elliptic operator, 
$\omega=\sqrt{H}$, and $f(z)$ be a function which is analytic in the right
half-plane decreasing at infinity in the sector $|\arg(z)|<\frac{\pi}{4}$.
We are trying to find a reduction formula
\be
\Tr f(\omega)=\int_0^\infty dt\; h(t)\Tr\exp(-tH),
\ee
with some function $h$, that reduces the calculation of the trace of $f(\omega)$
to the calculation of the trace of $\exp(-tH)$, that is, to the
classical heat trace.

It is not difficult to see that the function $h(t)$ is given by
the inverse Laplace 
transform of the function $f(\sqrt{z})$
\be
h(t)=\frac{1}{\pi i}\int_C dz\;z e^{tz^2}f(z);
\ee
here $C$ is a $<$-shaped contour in the right half-plane going from
$e^{-i\pi/4}\infty$ to the point $\varepsilon>0$ on the real axis and 
then to $e^{i\pi/4}\infty$ so that all poles of the integrand lie 
to the left of the contour $C$.
We consider the class of functions $f$ such that this integral converges.
Then one can show that the function $f$ can be represented by the integral
\be
f(x)=\int_0^\infty dt\; h(t) e^{-tx^2}.
\ee

By applying this method to the exponential function we get,
in particular,
\be
e^{-x}=(4\pi)^{-1/2}\int\limits_0^\infty
dt\;t^{-3/2}\exp\left(-\frac{1}{4t}\right)e^{-tx^2},
\label{44cca}
\ee
which is valid for any $x\ge 0$,

The functions introduced above can be represented as series
which converge for $x>0$,
\bea
E_f(x) &=& \sum_{k=1}^\infty (-1)^{k+1} e^{-kx},
\label{110xxc}
\\
E_b(x) &=& \sum_{k=1}^\infty e^{-kx},
\label{111xxc}
\\
E_0(x) &=& \sum_{k=0}^\infty e^{-(2k+1) x}.
\label{110xxcz}
\eea
By using the integral (\ref{44cca})
we find the integral representation
of these functions
\bea
E_{b,f,0}(x) &=& \int\limits_0^\infty dt\; h_{b,f,0}(t)\exp\left(-t x^2\right),
\label{311xx}
\eea
where 
\bea
h_f(t) &=& (4\pi)^{-1/2}t^{-3/2}
\sum_{k=1}^\infty (-1)^{k+1} k\exp\left(-\frac{k^2}{4t}\right),
\\
h_b(t) &=& (4\pi)^{-1/2}t^{-3/2}
\sum_{k=1}^\infty k\exp\left(-\frac{k^2}{4t}\right),
\\
h_0(t) &=& (4\pi)^{-1/2}t^{-3/2}
\sum_{k=0}^\infty \left( 2k+1\right)
\exp\left(-\frac{\left(2k+1\right)^2}{4t}\right).
\eea
By computing the inverse Laplace transform
these functions can also be written as 
\bea
h_{f}(t) &=& \frac{1}{2\pi }\fint_\RR dp\; p
\tan\left(\frac{p}{2}\right)\exp(-tp^2),
\\
h_{b}(t) &=& \frac{1}{2\pi }\fint_\RR dp\;p
\cot\left(\frac{p}{2}\right)\exp(-tp^2),
\\
h_{0}(t) &=& \frac{1}{2\pi }\fint_\RR dp\;
\frac{p}{\sin p}\exp(-tp^2),
\eea
where the
integrals are 
taken in the principal value sense; the imaginary part cancels out.
Obviously, we have
\be
h_0(t)=\frac{1}{2}\left[h_b(t)+h_f(t)\right].
\ee

All the functions $h_{b,f,0}(t)$ are exponentially small as $t\to 0$
\bea
h_{b,f,0}(t) \sim 
(4\pi)^{-1/2}t^{-3/2}
\exp\left(-\frac{1}{4t}\right).
\eea
As $t\to \infty$ these functions have the asymptotic expansions
\bea
h_{f}(t) &\sim&(4\pi)^{-1/2}
\sum_{k=1}^\infty (-1)^{k+1}\frac{(2^{2k}-1)B_{2k}}{2^{2k}k!}t^{-k-1/2},
\\[10pt]
h_{b}(t) &\sim& \frac{1}{\sqrt{\pi}}t^{-1/2}
-(4\pi)^{-1/2}\sum_{k=1}^\infty(-1)^{k+1}
\frac{B_{2k}}{2^{2k-1}k!}
t^{-k-1/2},
\eea
where $B_k$ are Bernoulli numbers. The leading terms of the asymptotics are 
\bea
h_f(t) &\sim&  \frac{1}{8\sqrt{\pi}}t^{-3/2},
\label{519viax}
\\
h_b(t) &\sim& \frac{1}{\sqrt{\pi}}t^{-1/2},
\label{520viax}
\\
h_0(t)&\sim& \frac{1}{2\sqrt{\pi}}t^{-1/2}.
\label{521viax}
\eea

\subsection{Heat Trace Representation}

Now, by using these integral representations
of the functions
$E_{b,f,0}$ we obtain the heat trace
representation for the Bogolyubov invariant. For the bosonic case
we get from (\ref{311xxa})
\be
B_b(\beta) = 
\int\limits_0^\infty dt\int\limits_0^\infty ds\;
h_f\left(t\right)
h_b\left(s\right)
\exp\left(-m^2\beta^2(s+t)\right)\Psi\left(\beta^2t,\beta^2s\right),
\label{413xxa}
\ee
where  $\Psi(s,t)$ is the function defined in (\ref{220xxa}).
This is the regularized version of the eq. (\ref{238xxa}).
For the fermionic case we get from (\ref{318xxa}) 
\bea
B_f(\beta) &=&
\int\limits_0^\infty dt\int\limits_0^\infty ds\;\,
h_0\left(s\right)
h_0\left(t\right)
\exp\left(-m^2\beta^2(s+t)\right)
\nonumber\\
&&\times
\beta^2\left\{\Phi\left(\beta^2t,\beta^2s\right)
+m^2\Psi\left(\beta^2t,\beta^2s\right)\right\},
\label{414xxa}
\eea
where $\Phi(t,s)$ is the function defined in (\ref{248xxa}).
This is the regularized version of the eq. (\ref{221xxb}).
It is easy to see that  the integrals
for the Bogolyubov invariant converge both as $t,s\to 0$
and (for sufficiently large $m^2$) also as $t,s\to \infty$.


We can also define the corresponding 
{\it Bogolyubov zeta function}
\bea
Z_{b,f}(s) &=&
\frac{1}{\Gamma(s)}\int_0^\infty d\beta \; \beta^{s-1}
B_{b,f}(\beta).
\eea
As will be shown below, as $\beta\to 0$ the Bogolyubov invariants behave as
$\beta^{-n}$, with $n=\dim M$, 
and, therefore, the zeta functions are analytic for
$\Real s>n$.
Thus, by using the inverse Mellin
transform we obtain the Bogolyubov invariant in terms of the zeta functions
\bea
B_{b,f}(\beta)=\frac{1}{2\pi i}\int_{c-i\infty}^{c+i\infty} ds\;
\beta^{-s}\Gamma(s)Z_{b,f}(s)\,,
\label{330bbb}
\eea
where $c>n$. 


Thus, we reduced the calculation of the Bogolyubov invariant to the 
calculation of the classical heat trace and the combined heat traces
studied in \cite{avramidi19a}
\bea
\Theta_{\pm}(t)&=&\Tr\exp(-tH_\pm),
\label{421xxc}
\\
X(t,s)&=&\Tr\exp(-tH_+)\exp(-sH_-),
\label{422xxc}
\\
Y(t,s)&=& \Tr A_+\exp(-tH_+)A_-\exp(-sH_-).
\label{423xxc}
\eea
Here $\Theta(t)$ is the standard classical heat trace; it has been studied a lot
in the literature (see, e.g. \cite{gilkey95,avramidi91,avramidi00,avramidi10}). 
The other traces are new.

For the fermionic case
we will find it useful to introduce more general traces
(recall that $H_\pm=A_\pm^2$)
\bea
\Xi_\pm(t,\alpha) &=& \Tr \exp(-t H_\pm+i\alpha A_\pm),
\label{424zzc}
\\
W(t,s;\alpha,\beta) &=& \Tr\exp(-tH_++i\alpha A_\pm)\exp(-sH_-+i\beta A_-).
\label{51xxa}
\eea
Then, obviously,
\bea
\Theta_\pm(t) &=& \Xi_\pm(t,0),
\\
X(t,s) &=& W(t,s;0,0),
\\
Y(t,s) &=& -\frac{\partial}{\partial \alpha}
\frac{\partial}{\partial \beta} W(t,s;\alpha,\beta)
\Big|_{\alpha=\beta=0}.
\eea
Therefore, all traces can be obtained from the traces 
(\ref{424zzc}) and (\ref{51xxa}).

Notice that the trace $\Xi(t,\alpha)$ satisfies the one-dimensional
heat equation on $\RR$,
\be
\partial_t \Xi_\pm(t,\alpha)=\partial_\alpha^2 \Xi_\pm(t,\alpha),
\ee
and, therefore, it can be written in the form
\be
\Xi_\pm(t,\alpha)=(4\pi t)^{-1/2}\int\limits_\RR d\alpha' 
\exp\left\{-\frac{(\alpha-\alpha')^2}{4t}\right\}
T_\pm(\alpha'),
\ee
where
\be
T_\pm(\alpha)=\Tr\exp\left(i\alpha A_\pm\right),
\ee
which should be understood in the distributional sense.
Similarly, the invariant $W(t,s;\alpha,\beta)$ satisfies the heat equations
\bea
\partial_t W(t,s;\alpha,\beta) &=& \partial_\alpha^2 W(t,s;\alpha,\beta),
\\
\partial_s W(t,s;\alpha,\beta) &=& \partial_\beta^2 W(t,s;\alpha,\beta),
\eea
and, therefore, it can be written in the form
\be
W(t,s;\alpha,\beta)=(4\pi)^{-1} (ts)^{-1/2}\int\limits_{\RR^2} d\alpha'd\beta' 
\exp\left\{-\frac{(\alpha-\alpha')^2}{4t}
-\frac{(\beta-\beta')^2}{4s}\right\}
S_\pm(\alpha',\beta'),
\ee
where
\be
S_\pm(\alpha,\beta)=\Tr\exp\left(i\alpha A_+\right)
\exp\left(i\beta A_-\right).
\ee

Notice also that for equal operators $H_-=H_+$ and $A_-=A_+$ the combined heat traces
can be expressed in terms of the classical one
\bea
X(t,s) &=&\Theta(t+s),
\label{57zzc}
\\
Y(t,s)&=& -\partial_t\Theta(t+s),
\label{58zzc}
\eea
also,
\be
W(t,s;\alpha,\beta)=\Xi(t+s,\alpha+\beta).
\ee
More complicated similar relations can be obtained by considering
distinct but commuting operators. For example, if the operators
differ by just a constant, $M^2$,
\be
H_+=H_-+M^2,
\ee
then
\be
X(t,s)=e^{-tM^2}\Theta_-(t+s).
\ee
Also, if the operators $A_+$ and $A_-$ differ by a constant
\be
A_+=A_-+M,
\ee
then
\be
W(t,s;\alpha,\beta)=e^{-tM^2+i\alpha M}\Xi_{-}(t+s,\alpha+\beta+2itM).
\ee

\subsection{Spectral Representation of Heat Traces}

Let $\{\lambda_k^\pm\}$ be the eigenvalues and $\{\varphi_k^\pm\}$ be the 
orthonormal sequence of eigensections
of the operator $H_\pm$. 
Then the heat traces (\ref{421xxc})-(\ref{423xxc}) take the following form
\bea
\Theta_{\pm}(t)
&=&\sum_{k=1}^\infty \exp\left(-t\lambda^\pm_k\right),
\label{532xxc}
\\
X(t,s)&=&
\sum_{k,j=1}^\infty \exp\left(-t\lambda^+_k-s\lambda^-_j\right)
\left|(\varphi^-_j,\varphi^{+}_k)\right|^2.
\label{533xxc}
\eea
Also, let $\mu_k^\pm$ be the eigenvalues of the operator $A_\pm$ 
(recall that
$H_\pm=A_\pm^2$).
Then
\bea
Y(t,s)&=& 
\sum_{k,j=1}^\infty \exp\left[-t(\mu^+_k)^2-s(\mu^-_j)^2\right]
\mu^+_k\mu^-_j\left|(\varphi^-_j,\varphi^{+}_k)\right|^2,
\label{534xxc}
\eea
and the generalized traces (\ref{424zzc}), (\ref{51xxa}), have the form
\bea
\Xi_\pm(t,\alpha)&=& 
\sum_{k=1}^\infty \exp\left[-t(\mu^\pm_k)^2+i\alpha\mu^\pm_k\right],
\label{56zza}
\\
W(t,s;\alpha,\beta)
&=&
\sum_{k,j=1}^\infty \exp\left[-t(\mu^+_k)^2+i\alpha\mu^+_k
-s(\mu^-_j)^2+i\beta\mu_j^-\right]
\left|(\varphi^-_j,\varphi^{+}_k)\right|^2.
\label{534xxd}
\eea

\subsection{Integral Representation of Heat Traces}

Laplace type operators $H_\pm$ naturally define
Riemannian metrics
$g_\pm$ and connections $\nabla^\pm$ on the vector bundle.
We use these metrics and connections to define the geodesic distance, the parallel transport,
the covariant derivatives etc.
Since while working with two different operators we do not have
a single metric, then, following \cite{avramidi04,avramidi19a},
we prefer to work with
the vector bundle of densities of weight $1/2$ and with the Lebesgue measure $dx$
instead of the Riemannian volume element $d\vol_g=dx\; g^{1/2}$,
with $g=\det(g_{ij})$.
Then the heat kernels  $U_\pm(t;x,x')$ of the heat semigroup $\exp(-tH_\pm)$
are also densities of weight $1/2$ at each point $x$ and $x'$, and the
diagonals $U_\pm(t;x,x)$ are densities of weight $1$.

The heat kernel of the operator $H_\pm$
has the following spectral representation
\be
U_\pm(t;x,x')=\sum_{k=1}^\infty \exp\left(-t\lambda^\pm_k\right)
\varphi^\pm_k(x)\varphi^{\pm*}_k(x').
\ee
Then the heat traces (\ref{421xxc})-(\ref{423xxc}) take the following form
\bea
\Theta_{\pm}(t)&=&\int\limits_M dx\;\tr U_\pm(t;x,x),
\label{532xxca}
\\
X(t,s)&=&\int\limits_{M\times M}dx\;dx'\;\tr U_+(t;x,x')U_-(s;x',x).
\label{533xxca}
\eea
Here and everywhere below $\tr$ denotes the fiber trace.

The integral kernel of the heat semigroup
$\exp(-tH_\pm+i\alpha A_\pm)$
has the form
\be
V_\pm(t,\alpha;x,x')
=\sum_{k=1}^\infty \exp\left[-t(\mu^\pm_k)^2+i\alpha\mu_k\right]
\varphi^\pm_k(x)\varphi^{\pm*}_k(x').
\ee
Therefore, (recall that $\lambda^\pm_k=(\mu^\pm_k)^2$)
\be
U_\pm(t;x,x')=V_\pm(t,0;x,x')
=\sum_{k=1}^\infty \exp\left[-t(\mu^\pm_k)^2\right]
\varphi^\pm_k(x)\varphi^{\pm*}_k(x').
\ee
Then
\be
A_\pm U_\pm(t;x,x')=\sum_{k=1}^\infty \exp\left(-t(\mu^\pm_k)^2\right)
\mu^\pm_k\varphi^\pm_k(x)\varphi^{\pm*}_k(x').
\ee
and
\bea
Y(t,s)&=& \int\limits_{M\times M}dx\;dx'\;\tr A_+U_+(t;x,x')A_-U_-(s;x',x),
\label{534xxca}
\eea
where the differential operators act on the first spacial
argument of the heat kernel. 

The generalized traces (\ref{424zzc}), (\ref{51xxa}), have the form
\bea
\Xi_\pm(t,\alpha)&=& 
\int\limits_M dx\;\tr V_\pm(t,\alpha;x,x),
\label{56zzaa}
\\
W(t,s;\alpha,\beta)
&=&\int\limits_{M\times M}dx\;dx'\tr V_+(t,\alpha;x,x')V_-(s,\beta;x',x).
\label{534xxda}
\eea

\section{Asymptotics of Heat Traces}
\setcounter{equation}0
\label{sec5}
\subsection{Heat Kernel Asymptotics of Laplace Type Operators}

First of all, it is easy to see that 
the asymptotics of the heat trace as $\beta\to \infty$ 
are determined by the bottom eigenvalues
\be
\Theta_{\pm}(\beta^2t) \sim \exp\left(-\beta^2t\lambda_1^\pm\right).
\ee
We will be primarily interested in the asymptotics as $\beta\to 0$.

For Laplace type operators $H_\pm$ 
there is an asymptotic expansion of the heat kernel
$U_\pm(t;x,x')$ 
in the neighborhood of the diagonal 
as $t\to 0$
(see e.g. \cite{avramidi91,avramidi00,avramidi10,avramidi15})
\bea
U_\pm(t;x,x')\sim 
(4\pi)^{-n/2}\exp\left(-\frac{\sigma_\pm}{2t}\right)
\sum_{k=0}^\infty  t^{k-n/2} \tilde a^\pm_k,
\label{513xxcd}
\eea
where $\sigma_\pm=\sigma_\pm(x,x')$ 
is the   Ruse-Synge function (also called the world function)
of the metric $g_\pm$ and
\bea
\tilde a^\pm_k
=\frac{(-1)^k}{k!}D_\pm^{1/2}\cP_\pm
a^\pm_k,
\label{513xxcq}
\eea
where 
$D_\pm=D_\pm(x,x')$ is the Van Vleck-Morette determinant,
$\cP_\pm=\cP_\pm(x,x')$ is the operator of parallel transport of sections along 
the geodesic in the connection $\nabla^\pm$ and the metric $g_\pm$
from the point $x'$ to the point $x$ and $a^\pm_k=a^\pm_k(x,x')$ are 
the usual heat kernel coefficients, in particular,
\cite{avramidi00}
\bea
[a_0] &=& I.
\eea

Therefore, there is the asymptotic expansion
of the classical heat trace (\ref{532xxc})
as \mbox{$\beta\to 0$,}
\be
\Theta_\pm\left(\beta^2 t\right) 
\sim (4\pi)^{-n/2}
\sum_{m=0}^\infty \beta^{2m-n} t^{m-n/2}A^\pm_m.
\label{513xxc}
\ee
where
\be
A^\pm_m=\frac{(-1)^m}{m!}\int\limits_M dx\; g_\pm^{1/2}\tr\; [a^\pm_m].
\ee
are the well known global heat trace coefficients
for the operators $H_\pm$ (notice the different normalization
factor compared to our earlier work
\cite{avramidi91,avramidi00,avramidi10,avramidi15}).
This is the classical heat trace asymptotics
of Laplace type operators.

\subsection{Combined Heat Trace Asymptotics}

It is easy to see again that the asymptotics of the combined heat traces
as $\beta\to \infty$ are 
determined by the bottom eigenvalues
\bea
X\left(\beta^2t,\beta^2s\right)
&\sim & \exp\left[-\beta^2\left(t\lambda_1^++s\lambda_1^-\right)\right]
\;|(\varphi^-_1,\varphi^+_1)|^2,
\\
Y\left(\beta^2t,\beta^2s\right)
&\sim & \exp\left[-\beta^2
\left(t(\mu_1^+)^2+s(\mu_1^-)^2\right)\right]\;\mu_1^+\mu_1^-\;
|(\varphi^-_1,\varphi^+_1)|^2.
\eea

We will be interested mainly in the asymptotics as $\beta\to 0$.
In \cite{avramidi19a} we proved the following theorem.
\begin{theorem}
\label{theorem1}
There are asymptotic expansions as $\beta\to 0$
\bea
X(\beta^2{}t,\beta^2{}s) &\sim& (4\pi)^{-n/2}
\sum_{k=0}^\infty \beta^{2k-n} B_k(t,s),
\label{1zaab}
\\
Y(\beta^2{}t,\beta^2{}s) &\sim&
(4\pi)^{-n/2}
\sum_{k=0}^\infty \beta^{2k-2-n} 
C_k(t,s),
\label{15zaac}
\eea
where 
$B_k(t,s)$ are homogeneous functions of $t$ and $s$ of degree
$(k-n/2)$ and $C_k(t,s)$ are homogeneous functions of $t$ and $s$ of degree
$(k-1-n/2)$. They are integrals of some scalar densities 
built polynomially from the symbols of the operators $H_\pm$ and $A_\pm$.

\end{theorem}
The coefficients $B_0$, $B_1$, $C_0$, and $C_1$ are computed explicitly
in \cite{avramidi19a}.

This enabled us also to compute the asymptotic expansion of the
relative spectral invariants as $\beta\to 0$;
they have the form \cite{avramidi19a} 
\bea
\Psi(\beta^2{}t,\beta^2{}s) 
&\sim& (4\pi)^{-n/2}
\sum_{m=0}^\infty \beta^{2k-n} \Psi_k(t,s),
\label{120via}
\\
\Phi(\beta^2{}t,\beta^2{}s) 
&\sim&
(4\pi)^{-n/2}
\sum_{k=0}^\infty \beta^{2k-2-n} 
\Phi_k(t,s),
\label{121via}
\eea
where
\bea
\Psi_k(t,s) &=&
(t+s)^{k-n/2}(A_k^++A_k^-)
-B_k(t,s)-B_k(s,t),
\label{543xxca}
\\
\Phi_k(t,s) &=&
-\left(k-\frac{n}{2}\right)(t+s)^{k-1-n/2}\left(A_k^++A_k^-\right)
-C_k(t,s)-C_k(s,t).
\label{542saax}
\eea

Notice that since for equal operators $H_-=H_+$ the 
combined trace $X(t,s)$
can be expressed in terms of the classical heat trace
(\ref{57zzc}),
then, by comparing (\ref{1zaab}) and 
(\ref{513xxc}) we see that in this case
\be
B_k(t,s)=(t+s)^{k-n/2} A_k.
\ee
This gives non-trivial relations between the heat kernel coefficients
and their derivatives and provides a useful check of the results.


Also, since for equal operators $A_-=A_+$ the 
combined trace $Y(t,s)$
can be expressed via
(\ref{58zzc})
in terms of the classical heat trace
then, by comparing (\ref{15zaac}) and 
(\ref{513xxc}) we see that in this case
\be
C_k(t,s)=-\left(k-\frac{n}{2}\right)(t+s)^{k-1-n/2} A_k.
\ee
This gives non-trivial relations between the heat kernel coefficients
and their derivatives and provides a useful check of the results.

\section{Asymptotics of Bogolyubov Invariant}
\setcounter{equation}0
\label{sec6}
\subsection{Mellin Transforms}
 
We use the Mellin transform to study the asymptotic expansion of the
integrals folowing \cite{avramidi91,avramidi00,avramidi10,avramidi17}.
Let $f$ be a smooth function on $\RR_+$.
Suppose that:
\begin{enumerate}
\item
it decreases at infinity faster than any power of $t$, that is,
\be
\lim_{t\to \infty}t^\gamma\partial_t^N f(t)=0
\label{71qqqa}
\ee
for any positive constant $\gamma>0$ and any non-negative intger $N\ge 0$, and
\item
there is a constant $\mu$ such that
\be
\lim_{t\to 0}t^\gamma\partial_t^N \left[t^\mu f(t)\right]=0
\label{71qqq}
\ee
for any positive constant $\gamma>0$ and
any non-negative integer
$N$.
\end{enumerate}

We consider a slightly modified version of the Mellin transform of the
function $f$ introduced in \cite{avramidi91}
\be
\hat f_q=\frac{1}{\Gamma(-q)}
\int\limits_0^\infty dt\;
t^{-q-1+\mu}f(t).
\label{72qqq}
\ee
The integral (\ref{72qqq}) converges for ${\rm Re}\, q<0$. By integrating by
parts $N$ times and using the asymptotic conditions (\ref{71qqq}) we also get
\be
\hat f_q
=\frac{1}{\Gamma(-q+N)}\int\limits_0^\infty dt\;
t^{-q-1+N}(-\partial_t)^N
\left[t^\mu f(t)\right].
\ee
This integral converges for ${\rm Re}\,q<N$ and, therefore,
defines an entire function. 
Now, by inverting the Mellin transform we obtain a useful 
integral representation
\be
f(t)=\frac{1}{2\pi i}\int\limits_{c-i\infty}^{c+i\infty}dq\;
t^{q-\mu}\,\Gamma(-q)\hat f_q,
\label{76qqq}
\ee
where $c<0$. By moving the contour of integration to the right
we obtain the asymptotic expansion of the function $f$
as $t\to 0$,
\be
f(t)\sim \sum_{k=0}^\infty \frac{(-1)^k}{k!}t^{k-\mu}\hat f_k.
\label{76viax}
\ee

Now, let $h$ be another smooth function on $\RR_+$.
Suppose that:
\begin{enumerate}
\item
it decreases as $t\to 0$ faster than any power of $t$,
that is,
\be
\lim_{t\to 0}t^{-\gamma}\partial_t^N h(t)=0
\label{71aqqz}
\ee
with any positive $\gamma>0$ and any non-negative integer $N\ge 0$, and
\item
there is a positive constant $\nu >0$ such that
\be
\lim_{t\to\infty}t^{-\gamma}\partial_t^N \left[t^\nu  h(t)\right]
=0
\label{71aqq}
\ee
with any positive $\gamma>0$ and any non-negative integer $N\ge 0$.
\end{enumerate}

We define a modified Mellin transform of the function $h$ by
\be
\hat h_q=\frac{1}{\Gamma(-q)}\int_0^\infty dt\;  t^{q-1+\nu }h(t).
\ee
This integral converges and 
defines an analytic function of $q$ for ${\rm Re}\, q<0$.
By integration by parts we get  for any $N>0$,
\be
\hat h_q
=\frac{1}{\Gamma(-q+N)}\int\limits_0^\infty dt\;
t^{q-1+N}\partial_t^N
\left[t^\nu  h(t)\right];
\ee
this integral converges for ${\rm Re}\, q<N$
and defines 
the analytic continuation to an entire
function.
By inverting the Mellin transform we get an integral representation
of the function $h$
\be
h(t)=\frac{1}{2\pi i}\int\limits_{c-i\infty}^{c+i\infty}dq\;
t^{-q-\nu }\,\Gamma(-q)\hat h_q
\label{76qqqz}
\ee
where $c<0$. By moving the contour of integration to the right
we get the asymptotic expansion of the function $h$
as $t\to\infty$
\be
h(t)\sim\sum_{k=0}^\infty \frac{(-1)^k}{k!}t^{-k-\nu }\hat h_k.
\ee

\begin{lemma}
\label{lemma9}
Let $f$ and $h$ be the functions described above
and $m=|\mu+\nu-1|$. Then the
integral
\be
I(\varepsilon)=\int_0^\infty dt\; h(t) f(\varepsilon t),
\ee
has the following asymptotic expansion as $\varepsilon\to 0$:
\begin{enumerate}
\item 
If $\mu+\nu $ is not a integer
then
\be
I(\varepsilon)\sim
\sum_{k=0}^\infty \varepsilon^{k-\mu}c_k^{(1)}
+\sum_{k=0}^\infty \varepsilon^{k+\nu-1}c_k^{(2)},
\label{712saa}
\ee
where
\bea
c_k^{(1)} &=& 
\frac{(-1)^{k}}{k!}\Gamma(-k+\mu+\nu-1) \hat h_{k-\mu-\nu+1} \hat f_k,
\\
c_k^{(2)} &=& 
\frac{(-1)^{k}}{k!}\Gamma(-k-\mu-\nu+1) \hat h_{k} \hat f_{k+\mu+\nu-1}. 
\eea

\item
If $\mu+\nu=1+m\ge 1$ is a positive integer with 
$m\ge 0$,
then
\be
I(\varepsilon)\sim
\sum_{k=0}^{m-1} \varepsilon^{k-\mu}c_k^{(3)}
+\sum_{k=0}^\infty \varepsilon^{k+m-\mu}c_k^{(4)}
+\log\varepsilon\sum_{k=0}^\infty \varepsilon^{k+m-\mu}c_k^{(5)},
\label{715saa}
\ee
where
\bea
c_k^{(3)} &=& 
\frac{(-1)^k}{k!}
(m-k-1)!H_{k-m}F_k,
\\
c_k^{(4)} &=& 
\frac{(-1)^m}{(k+m)!k!}\Bigl\{\left[\psi(k+1)+\psi(k+1+m)\right]
\hat h_k \hat f_{k+m}
\nonumber\\
&&
- \hat h'_k \hat f_{k+m}
-\hat h_k \hat f'_{k+m}
\Bigr\},
\\
c_k^{(5)} &=& 
-\frac{(-1)^m}{(k+m)!k!} \hat h_k \hat f_{k+m},
\eea
where $\psi(z)=\Gamma'(z)/\Gamma(z)$ 
is the logarithmic derivative of the gamma-function,
$\hat f'_k=\partial_q \hat f_q\big|_{q=k}$ and $\hat h'_k=\partial_q \hat h_q\big|_{q=k}$.
If $m=0$ then the first sum is absent in (\ref{715saa}).

\item
If $\mu+\nu=1-m\ge 0$ is a non-positive integer with
$m\ge 1$,
then 
\be
I(\varepsilon)\sim
\sum_{k=0}^{m-1} \varepsilon^{k-m-\mu}c_k^{(6)}
+\sum_{k=0}^\infty \varepsilon^{k-\mu}c_k^{(7)}
+\log\varepsilon\sum_{k=0}^\infty \varepsilon^{k-\mu}c_k^{(8)},
\label{719saa}
\ee
where
\bea
c_k^{(6)} &=& 
\frac{(-1)^{k}}{k!}
(m-k-1)! \hat h_{k} \hat f_{k-m},
\\
c_k^{(7)} &=& 
\frac{(-1)^m}{(k+m)!k!}\Bigl\{\left[\psi(k+1)+\psi(k+1+m)\right]
\hat h_{k+m} \hat f_{k}
\nonumber\\
&&
-\hat h'_{k+m} \hat f_{k} - \hat h_{k+m} \hat f'_{k}
\Bigr\},
\\
c_k^{(8)} &=& 
-\frac{(-1)^m}{(k+m)!k!} \hat h_{k+m} \hat f_{k}.
\eea

\end{enumerate}

\end{lemma}

\noindent
{\it Proof.}
By using the Mellin representations of the functions $f$ and $h$ we get
\be
I(\varepsilon)=
\frac{1}{2\pi i}\int\limits_{c-i\infty}^{c+i\infty}dq\;
\Gamma(-q) \Gamma(-q+\mu+\nu -1) \hat h_{q-\mu-\nu +1} \hat f_q \varepsilon^{q-\mu}
\ee
with $c$ being a sufficiently large negative constant such that the arguments of both
gamma functions have positive real parts, that is, 
$c<\min\{0, \mu+\nu -1\}$; then all singularities of the integrand
lie to the right of the contour of integration.
Integrals of this type are a particular case of the so-called Mellin-Barnes
integrals. They are a very powerful tool in computing the heat trace asymptotics.
(see Lemma 1 in 
\cite{avramidi17}).

Let
\be
\varphi(q)=\Gamma(-q) \Gamma(-q+\mu+\nu -1) \hat h_{q-\mu-\nu +1} \hat f_q
\ee
There are three essentially different cases.

Case I. In the case when the number $\mu+\nu $ is not a integer
the function $\varphi$ is meromorphic with
simple poles at the points $q=k$, $k=0,1,2,\dots$ and at the points
$q=k+\mu+\nu-1$, $k=0,1,2,\dots$. By moving the contour to the right
and using the well known analytic structure of the gamma function
\be
\Gamma(-k-z)=\frac{(-1)^k}{k!}\left\{
-\frac{1}{z}+\psi(k+1)+O(z)\right\},
\label{722saa}
\ee
where $\psi(z)=\Gamma'(z)/\Gamma(z)$, 
to evaluate the residues we obtain (\ref{712saa}).

Case II: 
Suppose that the number $\mu+\nu\ge 1$ is a positive integer,
that is, $\nu=-\mu+1+m$ with some non-negative
integer $m=\mu+\nu-1\ge 0$.
In this case
the function $\varphi$ is meromorphic with 
simple poles at the points $q=0,1,2,\dots, m-1$ 
(of course, if $m=0$ then there are no simple poles)
and double poles
at the points $q=k+m$, $k=0,1,2,\dots$.
By evaluating the residues we obtain (\ref{715saa}).

Case III: Suppose that the number $\mu+\nu\le 0$ is a non-positive integer, 
that is, that is, $\nu=-\mu+1-m$ with some positive
integer $m=-\mu-\nu+1\ge 1$.
Then the function $\varphi$ is meromorphic with 
simple poles at the points $q=-m, -m+1,\dots, -1$ and double poles
at the points $q=k$, $k=0,1,2,\dots$.
By evaluating the residues we obtain (\ref{719saa}).

\subsection{Laplace Type Operators}

Now we can compute the asymptotics of Bogolyubov invariant as $\beta\to 0$.
By introducing the integration variables 
\be
\rho=t+s, \qquad
u=\frac{t}{t+s},
\ee
we can write
the Bogolyubov invariant (\ref{413xxa}) in the form
\be
B_b(\beta) = 
\int_0^1 du\int\limits_0^\infty d\rho h(\rho,u) \psi(\beta^2\rho, u)
\ee
where
\bea
h(\rho,u) &=& \rho h_f\left(\rho u\right)h_b\left(\rho(1-u)\right),
\\
\psi(\rho, u) &=& \exp\left(-m^2\rho\right)
\Psi\left(\rho u, \rho (1-u)\right).
\eea
Now, we can apply Lemma \ref{lemma9} to compute the asymptotics
as $\beta\to 0$. 
By using the asymptotics of the function $\Psi(t,s)$, (\ref{120via}),
and the functions $h_f, h_b$,  (\ref{519viax}), (\ref{520viax}), 
it is easy to see that the functions $\psi$ and $h$ above satisfy all the conditions
of the lemma with $\mu=n/2$
and $\nu=1$. 
We define the Mellin transforms of the functions $h$ and $\psi$ by
\bea
\hat \psi_q(u) &=& \frac{1}{\Gamma(-q)}
\int\limits_0^\infty d\rho\;
\rho^{-q-1+n/2}\psi(\rho,u),
\label{72qqqza}
\\
\hat h_q(u)&=&\frac{1}{\Gamma(-q)}\int_0^\infty d\rho\;  \rho^{q} h(\rho,u).
\label{733viaxz}
\eea
It is worth pointing out that the values of the Mellin transform
at non-negative integer points, $k\ge 0$, are determined by the coefficients
of the asymptotic expansion (\ref{120via})
of the relative spectral invariant
\be
\hat\psi_k(u)=(4\pi)^{-n/2}\sum_{j=0}^k(-1)^{j+k}\frac{k!}{j!}m^{2j}\Psi_{k-j}(u,1-u).
\label{734viax}
\ee
The values at non-integer points, as well as the values of the derivatives
at integer points, $\hat\psi'_k$, are determined by the global behavior of the
relative spectral invariant and are not locally computable. 

The coefficients $\Psi_0(t,s)$ and $\Psi_1(t,s)$ are computed
explicitly
in \cite{avramidi19a}.
In particular,
\be
\Psi_{0}(t,s)=(t+s)^{-n/2}\left(A_0^++A_0^-\right)
-B_0(t,s)-B_0(s,t),
\ee
where
\be
A_0^\pm=N\int_M dx\; g_\pm^{1/2},
\label{736viax}
\ee
with  $N=\tr I=\dim \cV$ and $g_\pm=(\det g_\pm^{ij})^{-1}$,
is the standard first heat kernel coefficient and 
\be
B_0(t,s)=N\int_M dx\; g^{1/2}(t,s);
\ee
here $g(t,s)=\det g_{ij}(t,s)$, and $g_{ij}(t,s)$ is the 
inverse of the matrix 
\be
g^{ij}(t,s)=tg_+^{ij}+sg_-^{ij}.
\label{738viax}
\ee

We have to distinguish the cases of even and odd dimension.\\
Case I. Odd dimension, $n=2m+1$. Then $\mu+\nu=m+3/2$ is not an integer and the asymptotics 
is given by (\ref{712saa}).
\be
B_b(\beta)\sim
\sum_{k=0}^\infty \beta^{2k-n}c_k^{(1)}
+\sum_{k=0}^\infty \beta^{2k}c_k^{(2)},
\label{712saaz}
\ee
where
\bea
c_k^{(1)} &=& 
\frac{(-1)^{k}}{k!}\Gamma(-k+n/2)\int_0^1 du\;\hat h_{k-n/2}(u)\hat\psi_k(u),
\\
c_k^{(2)} &=& 
\frac{(-1)^{k}}{k!}\Gamma(-k-n/2)
\int_0^1 du\;\hat h_{k}(u)\hat\psi_{k+n/2}(u). 
\eea
Notice that the coefficients $c^{(1)}_k$ of the all
odd powers of $\beta$
are locally computable invariants whereas the coefficients
$c^{(2)}_k$ of the even non-negative powers of $\beta$ are non-locally 
computable global invariants.


Case II. Even dimension, $n=2m$. Then $\mu+\nu=m+1$ is an integer and the asymptotics 
is given by (\ref{715saa}).
\be
B_b(\beta)\sim
\sum_{k=0}^{m-1} \beta^{2k-n}c_k^{(3)}
+\sum_{k=0}^{\infty} \beta^{2k}c_k^{(4)}
+\log\beta^2\;\sum_{k=0}^\infty \beta^{2k}c_k^{(5)},
\label{715saz}
\ee
where
\bea
c_k^{(3)} &=& 
\frac{(-1)^k}{k!}
\Gamma(-k+n/2)
\int_0^1du\; \hat h_{k-n/2}(u)\hat\psi_k(u),
\\
c_k^{(4)} &=& 
\frac{(-1)^m}{(k+m)!k!}\int_0^1 du\;
\Bigl\{\left[\psi(k+1)+\psi(k+1+m)\right]
\hat h_k(u) \hat\psi_{k+m}(u)
\nonumber\\
&&
-\hat h'_k(u)\hat\psi_{k+m}(u)-\hat h_k(u)\hat\psi'_{k+m}(u)
\Bigr\},
\\
c_k^{(5)} &=& 
-\frac{(-1)^m}{(k+m)!k!}
\int_0^1du\; \hat h_k(u) \hat\psi_{k+m}(u).
\eea
Notice that the coefficients $c^{(3)}_k$ and $c^{(5)}_k$ 
of the singular part and the logarithmic part
are locally computable invariants whereas the coefficients
$c^{(4)}_k$ of the regular part are not.
Also, the coefficients $c^{(3)}$, when written for general $n$,
have the same form as the coefficients $c^{(1)}_k$.
Therefore, the singular part of the asymptotics
containing the negative powers of $\beta$ has the same form
in both cases, regardless where the dimension $n$ is even or odd.

The leading asymptotics have the form
\be
B_b(\beta)=\beta^{-n}c_0^{(1)}+\beta^{-n+2}c^{(1)}_1 +O(\beta^{-n+4})+O(\log\beta).
\ee
By using the Mellin transform (\ref{733viax}) of the function $h$
and changing the integration variables $(\rho,u)\mapsto (t,s)$
we can rewrite the coefficients $c^{(1)}_k$ in the form
\be
c_k^{(1)} = \frac{(-1)^k}{k!}
\int\limits_0^\infty dt\int\limits_0^\infty ds\;
(t+s)^{k-n/2}h_f(t)h_b(s)\hat\psi_k\left(\frac{t}{t+s}\right).
\ee
Now, by using (\ref{734viax}) and the homogeneity property of the coefficients
$\Psi_k(t,s)$ (they are homogeneous functions of $t$ and $s$ of degree
$k-n/2$) we obtain
\bea
c_0^{(1)} &=& (4\pi)^{-n/2}
\int\limits_0^\infty dt\int\limits_0^\infty ds\; h_f(t)h_b(s)\Psi_0(t,s),
\\
c_1^{(1)} &=& (4\pi)^{-n/2}
\int\limits_0^\infty dt\int\limits_0^\infty ds\; h_f(t)h_b(s)
\left\{\Psi_1(t,s)-m^2(t+s)\Psi_0(t,s)
\right\}.
\eea
One can show that 
the coefficient $c^{(1)}_0$ is nothing but the coefficient $V_b$ computed in 
(\ref{417viax}).

\subsection{Dirac Type Operators}

Following the same strategy we compute the asymptotics of the 
Bogolyubov invariant for the Dirac type operators.
We have
\be
B_f(\beta) = \beta^2
\int\limits_0^1 du\int\limits_0^\infty d\rho\;
\chi(\rho,u)\varphi(\beta^2\rho,u),
\ee
where
\bea
\chi(\rho,u)  &=& \rho h_0(\rho u)h_0(\rho(1-u)),
\\
\varphi(\rho,u) &=& \exp(-m^2\rho)
\left\{\Phi\left(\rho u, \rho(1-u)\right)
+m^2\Psi\left(\rho u, \rho (1-u)\right)\right\}.
\label{414xxz}
\eea

Now, we can apply Lemma \ref{lemma9} to compute the asymptotics
as $\beta\to 0$. 
By using the asymptotics of the functions $\Phi(t,s)$, (\ref{121via}), and
$\Psi(t,s)$, (\ref{120via}),
and the functions $h_f, h_b$,  (\ref{519viax}), (\ref{520viax}), 
it is easy to see that the functions $\varphi$ and $\chi$ above satisfy all the conditions
of the lemma with $\mu=n/2+1$
and $\nu=0$. 
We define the Mellin transforms of the functions $\chi$ and $h$ by
\bea
\hat\varphi_q(u) &=& \frac{1}{\Gamma(-q)}
\int\limits_0^\infty d\rho\;
\rho^{-q+n/2}\varphi(\rho,u),
\label{72qqqz}
\\
\hat\chi_q(u)&=&\frac{1}{\Gamma(-q)}\int_0^\infty d\rho\;  \rho^{q-1}\chi(\rho,u).
\label{733viax}
\eea
It is worth pointing out that the values of the Mellin transform
at non-negative integer points, $k\ge 0$, are determined by the coefficients
of the asymptotic expansion (\ref{120via})
of the relative spectral invariant
\be
\hat\varphi_0(u) = (4\pi)^{-n/2}\Phi_{0}(u,1-u)
\ee
and for $k\ge 1$
\bea
\hat\varphi_k(u) = (4\pi)^{-n/2}
\sum_{j=0}^{k}\frac{(-1)^{j+k}k!}{j!}m^{2j}\left\{\Phi_{k-j}(u,1-u)
-j\Psi_{k-j}(u,1-u)\right\}.
\label{734viaz}
\eea
The values at non-integer points, as well as the values of the derivatives
at integer points, $\hat\chi'_k$, are determined by the global behavior of the
relative spectral invariant and are not locally computable. 

The coefficients $\Phi_0(t,s)$ and $\Phi_1(t,s)$ are computed
explicitly
in \cite{avramidi19a}.
In particular,
\be
\Phi_{0}(t,s)=\frac{n}{2}(t+s)^{-n/2-1}\left(A_0^++A_0^-\right)
-C_0(t,s)-C_0(s,t),
\ee
where $A_0^\pm$ are the standard first heat kernel coefficients
(\ref{736viax}) and 
\be
C_0(t,s)=\int_M dx\; g^{1/2}(t,s)
\frac{1}{2}g_{ij}(t,s)\tr\,
\left(\gamma_+^{i}\gamma_-^{j}\right),
\ee
where $\gamma^i_\pm$ are Dirac matrices determined by the leading symbols
of the operators $A_\pm$, (\ref{425viax}).

We have again two cases.\\
Case I. Odd dimension, $n=2m+1$. Then $\mu+\nu=m+3/2$ is not an integer and the asymptotics 
is given by (\ref{712saa}).
\be
B_f(\beta)\sim
\sum_{k=0}^\infty \beta^{2k-n}d_k^{(1)}
+\sum_{k=0}^\infty \beta^{2k}d_k^{(2)},
\label{712saax}
\ee
where
\bea
d_k^{(1)} &=& 
\frac{(-1)^{k}}{k!}\Gamma(-k+n/2)\int_0^1 du\;\hat \chi_{k-n/2}(u)\hat\varphi_k(u),
\\
d_k^{(2)} &=& 
\frac{(-1)^{k}}{k!}\Gamma(-k-n/2)
\int_0^1 du\;\hat \chi_{k}(u)\hat\varphi_{k+n/2}(u). 
\eea
Notice that the coefficients $d^{(1)}_k$ of the all
odd powers of $\beta$
are locally computable invariants whereas the coefficients
$d^{(2)}_k$ of the even non-negative powers of $\beta$ are non-locally 
computable global invariants.


Case II. Even dimension, $n=2m$. Then $\mu+\nu=m+1$ is an integer and the asymptotics 
is given by (\ref{715saa}).
\be
B_f(\beta)\sim
\sum_{k=0}^{m-1} \beta^{2k-n}d_k^{(3)}
+\sum_{k=0}^{\infty} \beta^{2k}d_k^{(4)}
+\log\beta^2\;\sum_{k=0}^\infty \beta^{2k}d_k^{(5)},
\label{715sazd}
\ee
where
\bea
d_k^{(3)} &=& 
\frac{(-1)^k}{k!}
\Gamma(-k+n/2)
\int_0^1du\; \hat \chi_{k-n/2}(u)\hat\varphi_k(u),
\\
d_k^{(4)} &=& 
\frac{(-1)^m}{(k+m)!k!}\int_0^1 du\;
\Bigl\{\left[\psi(k+1)+\psi(k+1+m)\right]
\hat \chi_k(u) \hat\varphi_{k+m}(u)
\nonumber\\
&&
-\hat \chi'_k(u)\hat\varphi_{k+m}(u)
-\hat \chi_k(u)\hat\varphi'_{k+m}(u)
\Bigr\},
\\
d_k^{(5)} &=& 
-\frac{(-1)^m}{(k+m)!k!}
\int_0^1du\; \hat \chi_k(u) \hat\varphi_{k+m}(u).
\eea
Notice that the coefficients $d^{(3)}_k$ and $d^{(5)}_k$ 
of the singular part and the logarithmic part
are locally computable invariants whereas the coefficients
$d^{(4)}_k$ of the regular part are not.
Also, the coefficients $d^{(3)}$, when written for general $n$,
have the same form as the coefficients $d^{(1)}_k$.
Therefore, the singular part of the asymptotics
containing the negative powers of $\beta$ has the same form
in both cases, regardless whether the dimension $n$ is even or odd.

The leading asymptotics have the form
\be
B_f(\beta)=\beta^{-n}d_0^{(1)}+\beta^{-n+2}d^{(1)}_1 +O(\beta^{-n+4})+O(\log\beta).
\ee
By using the Mellin transform (\ref{733viax}) of the function $\chi$
and changing the integration variables $(\rho,u)\mapsto (t,s)$
we can rewrite the coefficients $d^{(1)}_k$ in the form
\be
d_k^{(1)} = \frac{(-1)^k}{k!}
\int\limits_0^\infty dt\int\limits_0^\infty ds\;
(t+s)^{k-1-n/2}h_0(t)h_0(s)\hat\varphi_k\left(\frac{t}{t+s}\right).
\ee
Now, by using (\ref{734viaz}) and the homogeneity property of the coefficients
$\Phi_k(t,s)$ (they are homogeneous functions of $t$ and $s$ of degree
$(k-1-n/2)$) and
$\Psi_k(t,s)$ (they are homogeneous functions of $t$ and $s$ of degree
$(k-n/2)$) we obtain
\bea
d_0^{(1)} &=& (4\pi)^{-n/2}
\int\limits_0^\infty dt\int\limits_0^\infty ds\; h_0(t)h_0(s)\Phi_0(t,s),
\\
d_1^{(1)} &=& (4\pi)^{-n/2}
\int\limits_0^\infty dt\int\limits_0^\infty ds\; h_0(t)h_0(s)
\left\{\Phi_1(t,s)+m^2\left[-(t+s)^{}\Phi_0(t,s)+\Psi_0(t,s)\right]
\right\}.
\nonumber\\
\eea
One can show that 
the coefficient $d^{(1)}_0$ is nothing but the coefficient $V_f$ computed in 
(\ref{429viax}).


\section{Solvable Cases}
\setcounter{equation}0
\label{sec7}
\subsection{Equal Operators}

First of all, we notice that since for equal operators $H_-=H_+$ the 
combined trace $X(t,s)$
can be expressed in terms of the classical heat trace
\be
X(t,s) =\Theta(t+s),
\ee
then, by comparing (\ref{1zaab}) and 
(\ref{513xxc}) we see that in this case
\be
B_k(t,s)=(t+s)^{k-n/2} A_k.
\label{674viax}
\ee
Similarly, since for equal operators $A_-=A_+$ the 
combined trace $Y(t,s)$
can be expressed
in terms of the classical heat trace
\be
Y(t,s) = - \partial_t\Theta(t+s),
\ee
then, by comparing (\ref{15zaac}) and 
(\ref{513xxc}) we see that in this case
\be
C_k(t,s)=-\left(k-\frac{n}{2}\right)(t+s)^{k-1-n/2} A_k.
\label{676viax}
\ee
It is easy to see then that 
for equal operators $L_-=L_+$ and $D_-=D_+$ 
the relative spectral invariants vanish, 
$\Psi(t,s)=\Phi(t,s)=0$
and therefore, the Bogolyubov invariant vanishes
\be
B_b(\beta)=B_f(\beta)=0.
\ee

\subsection{Constant Potential Term}

If the Laplace type operators
differ by just a constant,
\be
H_+=H_-+M^2,
\ee
then the metrics and the connections are the same and
\bea
\Theta_+(t) &=& e^{-tM^2}\Theta_-(t),
\\
X(t,s) &=& e^{-tM^2}\Theta_-(t+s),
\eea
and, therefore,
\be
\Psi(t,s)=\left(e^{-tM^2}-1\right)\left(e^{-sM^2}-1\right)\Theta_-(t+s).
\ee
In this case
\bea
B_0(t,s) &=& (t+s)^{-n/2}A_0^-,
\\
B_1(t,s) &=& (t+s)^{1-n/2}A_1^--t(t+s)^{-n/2}M^2A_0^-.
\eea

For the Dirac case suppose that there is an endomorphism $M$
such that it anticommutes with the operator
$A_-$,
\be
A_-M=-MA_-,
\ee
and $M^2$ is a scalar.
Then it is easy to see that
\be
\Tr MA_-\exp(-sA^2_-)=0.
\ee
Now, suppose that
\be
A_+=A_-+M,
\ee
so that (recall that $H_+=A_+^2$)
\be
H_+=H_-+M^2;
\ee
Then it is easy to show that 
\bea
Y(t,s) &=& -e^{-tM^2}\partial_t\Theta_-(t+s),
\eea
and, hence,
\be
\Phi(t,s)=-\left(e^{-tM^2}-1\right)\left(e^{-sM^2}-1\right)
\partial_t\Theta_-(t+s)
+M^2e^{-(t+s)M^2}\Theta_-(t+s).
\ee
Therefore,
\bea
C_0(t,s) &=& \frac{n}{2}(t+s)^{-1-n/2}A_0^-,
\\
C_1(t,s) &=& \left(\frac{n}{2}-1\right)(t+s)^{-n/2}A_1^-
-\frac{n}{2}t(t+s)^{-1-n/2}M^2A_0^-.
\eea


A more general case is the case of {\it commuting}
operators; then the combined heat traces still simplify significantly,
they can be expressed in terms of the classical one
\bea
X(t,s) &=& \Tr\exp(-tH_+-s H_-),
\label{238ccxa}
\\
Y(t,s) &=& \Tr A_-A_+\exp(-tH_+ -sH_-).
\eea
Therefore, the asymptotics of the combined traces can be obtained from the 
classical ones.
Notice that the leading symbol of the operators $H(t,s)=tH_++sH_-$ 
is determined exactly by
the metric $g^{ij}(t,s)$. Therefore,
in this case the combined traces are given by the
classical trace for the operator $H(t,s)$.

\subsection{Nilpotent Lie Algebra}

Now, suppose that there are two sets of operators $\nabla_i^+, \nabla^-_j$ 
forming the Lie algebra
\bea
[\nabla^+_i,\nabla^+_j] &=& \cR^+_{ij},
\\{}
[\nabla^-_i,\nabla^-_j] &=& \cR^-_{ij},
\\{}
[\nabla^+_i,\nabla^-_j] &=& \cR_{ij},
\label{843viaz}
\eea
where
\be
\cR_{ij}=\frac{1}{2}\left(\cR^+_{ij}+\cR^-_{ij}\right),
\ee
all other commutators being zero.
We define two operators
\bea
H^\pm &=& -g_\pm^{ij}(\nabla^\pm_i+B^\pm_i)(\nabla^\pm_j+B^\pm_j)+Q_\pm.
\eea
where $g^{ij}_\pm$ are constant positive matrices,
$B^\pm_i$ are constant vectors and $Q_\pm$
are some constants. 
Then one can prove the following theorem
for the heat semigroup \cite{avramidi93,avramidi15}.
\begin{theorem}
\label{theorem2viaz}
The heat semigroup $\exp(-tH_\pm)$ 
can be presented in form of an
average over the Lie group with the Gaussian measure
\be
\exp(-tH^\pm) = (4\pi)^{-n/2}\Omega_\pm(t)\exp(-tQ_\pm)
\int\limits_{\RR^n} d\xi
\exp\left\{-\frac{1}{4}\left<\xi,D_\pm(t)\xi\right>
+\left<B^\pm,\xi\right>\right\}
\exp\left<\xi,\nabla^\pm\right>\,.
\label{858viax}
\ee
where $D=(D_{ij})$
is the matrix defined by
\be
D(t)=\cR\coth(tg^{-1}\cR)\,
\ee
and
\be
\Omega(t)=\det \left(\frac{\sinh(tg^{-1}\cR)}{\cR}\right)^{-1/2}.
\ee
\end{theorem}
By using this representation one can compute the
heat semigroup convolution
\be
U(t,s)=\exp(-tH_+)\exp(-sH_-)
\ee
exactly. We are going to carry this out in a separate 
work.

\section{Conclusion}

The goal of this paper was to introduce and to study new 
spectral invariants of two elliptic operators on manifolds
that we call the Bogolyubov invariants, the bosonic one 
$B_b(\beta)$ and the fermionic one, $B_f(\beta)$,
which depend on an adiabatic parameter $\beta$.
We established the general asymptotic expansion of these invariants 
as $\beta\to 0$ in terms of the so-called
relative spectral invariants studied in our paper
\cite{avramidi19a} 
and computed the first two coefficients of the asymptotic expansions.


\end{document}